\documentclass[11pt]{article}
\usepackage{epsfig}
\usepackage{latexsym}
\usepackage{amsmath}

\newtheorem{theorem}{Theorem}[section]{\bf}{\it}
\newtheorem{definition}[theorem]{Definition}{\bf}{\rm}
\newtheorem{proposition}[theorem]{Proposition}{\bf}{\it}
{\bf}{\it}
{\bf}{\it}
\renewcommand{\forall}{\mbox{for all}\,\,}
          \def\dt{\cal}

          \def\dM{{\dt M}}

          \def\E{{\cal E}}

          \def\H{{\cal H}}

          \def\O{{\cal O}}

          \def\U{{\cal U}}

      \def\gO{\Omega}
          \def\go{\omega}

          \def\complex{{\bf C}}

          \def\Im{{\rm Im}\,}

          \def\naturals{{\bf N}}

          \def\Rd{\reals^{1+s}}
          \def\reals{{\bf R}}

\title{Moving Quantum Systems: Particles Versus Vacuum}
\author{Bernd Kuckert\\II. Institut f\"ur Theoretische Physik\\
Luruper Chaussee 149\\22761 Hamburg, Germany\\
e-mail: bernd.kuckert@desy.de}
\date{}

\begin{document}
\maketitle

\begin{abstract}
  We give an overview on a couple of recent results concerning the
  KMS-condition and the characterization of thermodynamic equilibrium
  states from a moving observer's point of view. These results include
  a characterization of vacuum states in relativistic 
  quantum field theory and a general derivation of the Unruh effect.
\end{abstract}

\section{Introduction}

The KMS-condition introduced by Kubo, Martin, and Schwinger in
\cite{Kubo, MarS} is a generalization of the Gibbs characterization of
thermodynamic equilibrium states which is widely used in physics
today. In contrast to the Gibbs condition, the KMS-condition is
meaningful also for any infinitely extended system, whose Hamiltonian
$H$ does, in general, not have the property that $e^{-\beta H}$ is a
trace class operator ($\beta$ denotes the inverse temperature).

The relevance of the KMS-condition for the operator algebraic
description of thermodynamic equilibrium states was first discussed by
Haag, Hugenholtz, and Winnink \cite{HHW,Haa92}.  A general axiomatic
derivation of the KMS-condition\footnote{For the sake of brevity, the
  case that $H\geq0$ is occasionally considered as the KMS-condition
  at zero temperature.} from first principles was first proposed by
Haag et al.  (see \cite{Haa92,Kas76} and references given there), who
showed that the KMS-condition follows from the dynamical stability of
the state against small perturbations of the Hamitonian, together with
some clustering properties exhibited by many examples. A bit later,
Pusz and Woronowicz proved the equivalence of the KMS-condition at
nonnegative temperature on the one hand and a condition called
complete passivity on the other \cite{PW78}, which can be derived from
the Zeroth, the First, and the Second Law of Thermodynamics without
making any more technical assumptions. A couple of years ago, Bros and
Buchholz introduced a relativistic KMS-condition in quantum field
theory, which, in contrast to the usual KMS-condition, is not confined
to a distinguished frame of reference \cite{BB94}. Their conjectures
and ideas on how to derive this condition from first principles of
thermodynamics are partially confirmed by the results to be discussed
below, and they have been an important motivation for this research.

The discussion to follow is a summary of Refs. \cite{Kuc01} and
\cite{Kuc01a}. Ref. \cite{Kuc01} also contains a more detailed
introduction into the notions, ideas, and further references.

The passivity property of thermodynamic equilibrium states
means that no cyclic process, i.e., no temporary perturbation
of the Hamiltonian by observables, can perform more work than
it requires in order to take place. This can be justified
by the Second Law of Thermodynamics, and the First Law of 
Thermodynamics can be used to justify the mathematical formulation 
of the condition. Complete passivity strengthens this condition in 
a fashion that can be justified by the Zeroth Law of Thermodynamics.

Already the passivity condition can hold in one frame of reference
at most whenever matter is present, as moving matter drives 
windmills, turbines, etc.. In \cite{Kuc01}, 
the Pusz-Woronowicz result has been generalized 
by replacing, on the one hand, the condition of complete
passivity by the weaker condition of {\it complete semipassivity}, 
and by replacing, on the other hand, the
KMS-condition by the requirement that the KMS-condition is
fulfilled in {\it some} inertial frame. Both conditions can hold in 
several frames of reference for translation invariant states of 
matter. This will be discussed in Sect. \ref{matter}.

Recently it was proved by Guido and Longo in \cite{GL01} that the
KMS-condition at a finite and nonnegative temperature $\beta$ is
equivalent to a condition called {\it complete
$\beta$-boundedness}. This condition imposes a bound on the number of
degrees of freedom in certain phase space regions and is a weak form
of the Buchholz-Wichmann nuclearity condition \cite{BuWi}, very
similar to the (weaker) Haag-Swieca compactness criterion \cite{HaSw}.
While these criteria have been suggested and investigated in the 
setting of algebraic quantum field theory, complete $\beta$-boundedness
characterizes thermodynamic equilibrium states at inverse temperature
$\beta$ in much more general terms and without referring to 
any field theoretic structures. The result by Guido and Longo 
suggests to look for a 
modification of the $\beta$-boundedness condition that can hold,
like semipassivity, in several frames of reference, and leads to the
same conclusions as obtained for semipassivity. Such a condition is
{\it semi-$\beta$-boundedness}; this result of \cite{Kuc01a} will be discussed
in Sect. \ref{semibeta}.

On the other hand, passivity can hold in several frames of reference
in the absence of matter. Taking this as a basic characterization of a
vacuum state in quantum field theory, one can prove the spectrum
condition and the Unruh effect, i.e., the Bisognano-Wichmann symmetries 
\cite{Unr76,BW75,BW76}, for all pure
vacuum states of a local quantum field theory. This result of
\cite{Kuc01} will be discussed in Sect.  \ref{vacuum}. Similar results
for de Sitter and Anti-de Sitter spacetimes have been obtained in
\cite{BB98} and \cite{BFS}, respectively. Recent results
concerning the Bisognano-Wichmann symmetries of algebraic quantum
field theories in Minkowski space can be found in
\cite{BDFS98,Kuc00,Mun01,Bor00} and the references given there,
and for results concerning the possible role of these symmetries
in Robertson-Walker spacetimes, see \cite{BMS01}.

In the Conclusion (Sect. \ref{conc}), a couple of remarks are made
concerning the notion of particle and the Unruh effect.

\section{Semipassivity and homogeneous states of matter}\label{matter}

Our setting is as follows:

\begin{itemize}
\item A von Neumann algebra $\dM$ on a Hilbert space $\H$ 
represents the observable quantities characterizing the 
system under consideration.

\item For $s\geq1$, a family of $1+s$ commuting self-adjoint
operators $(H,{\bf P})=(H,P_1,\dots,P_s)$ generates a (strongly 
continuous and 
unitary) representation $V$ of the spacetime translation
group $(\Rd,+)$. 

\item A $V$-invariant state $\go$ of $\dM$, our object of
investigation, is
induced by a cyclic vector $\gO$ of $\dM$, and
$H\Omega=P_1\Omega=\dots=P_s\Omega=0$.
\end{itemize}

Define $S_\beta:=\{z\in\complex:\,-\beta<\Im z<0\}$.
$\go$ is called a {\it KMS-state at the inverse temperature} $\beta\geq0$
if there exists a continuous function $F:\overline{S}_\beta\to\complex$
that is analytic in $S_\beta$ and satisfies
\begin{equation}\label{KMS}
F(t)=\go(B_t A)\quad\mbox{and}\quad
F(t-i\beta)=\go(AB_t)\quad\forall t\in\reals,\quad A,B\in\dM.
\end{equation}
A KMS-state at $\beta=0$ (infinite temperature) is a 
trace, i.e., $\go(AB)=\go(BA)$ for all
$A,B\in\dM$, and
$\go$ may be considered a ``KMS-state at zero temperature''
if $H\geq0$, i.e., if $\go$ is a {\it ground state} of $H$.

As explained in \cite{PW78}, a cyclic change
of the Hamiltonian is represented by a unitary operator 
in the unit operator's norm-connected component $\U_1(\dM)$ 
of the unitary group $\U(\dM)$ of $\dM$. If the system is
in the above state $\go$ before the cycle $W\in\U_1(\dM)$ is 
started, the total amount of
work performed on the system during the cycle $W$ is
$L_W:=\langle W\Omega,[H,W]\Omega\rangle=
\langle W\Omega,HW\Omega\rangle$.\footnote{Note that
$H$ is an unbounded operator and that $\langle W\Omega,HW\Omega\rangle$
does not need to be defined for all $W\in\U_1(\dM)$. The
expression $\langle Hx,Wy\rangle-\langle x,WHy\rangle$
is defined for all $x,y$ in the domain of
$H$. $[H,W]\in\dM$ means that the sesquilinear form defined this way
is bounded and that the associated bounded operator is an element of
$\dM$; commutators involving $H$ or ${\bf P}$ are to be read 
this way.}

The state $\omega$ is called {\it passive} if $-L_W\leq0$ for
all cycles $W\in\U_1(\dM)$ with $[H,W]\in\dM$. It is 
{\it completely passive} if for each $N\in\naturals$, 
the state $\omega^{\otimes N}$ of the $N$th tensorial
$\dM^{\bigotimes N}$ of $\dM$ defined by
$$\dM^{\bigotimes N}\ni A_1\otimes A_2\otimes\dots
\otimes A_N\mapsto\go(A_1)\go(A_2)\dots\go(A_n)$$
is passive. As shown
in \cite{PW78}, a state is completely passive if and only if it is
a KMS-state or a ground state.

Suppose that $\go$ is passive with respect to the Hamiltonian $H$.
If the system is not at rest, but moves at a velocity ${\bf u}$,
then the time evolution is not generated by $H$, but by
$H+{\bf u P}$.\footnote{In a relativistic theory, this
generator must be multiplied by the time dilation factor
$\gamma=(1-{\bf u}^2/c^2)^{-1/2}$; see below.}
If $W\in\U_1(\dM)$ satisfies $[H,W]\in\dM$ and $[{\bf P},W]\in\dM$,
then the work performed by the corresponding cycle is
\begin{align*}
-L&=-\langle W\gO,
(H+{\bf u}{\bf P})\,W\gO\rangle\\
&\leq-\langle W\gO,{\bf u}{\bf P}\,W\gO\rangle,
\end{align*}
as $\go$ has been assumed to be passive with respect to $H$.
Defining $|{\bf P}|:=\sqrt{P_1^2+\dots+P_s^2}$, one finds
$-{\bf u}{\bf P}\leq|{\bf u}|\,|{\bf P}|$,
so
\begin{equation}\label{presemip}
-L\leq |{\bf u}|\langle W\gO,|{\bf P}|\,W\gO\rangle.
\end{equation}

Now suppose that $\go$ is not necessarily passive with respect
to $H$. We call
$\go$ {\em semipassive} if the work a cycle can perform is
bounded as in Ineq. (\ref{presemip}), i.e., if
there is a constant $\E\geq0$ such that
\begin{equation}\label{semipassivity}
-\langle W\gO,H\,W\gO\rangle
\leq\E\langle W\gO,|{\bf P}|\,W\gO\rangle
\end{equation}
for all $W\in\U_1(\dM)$ with
$[H,W]\in\dM$ and $[{\bf P},W]\in\dM$. The constant $\E$
will be referred to as an {\em efficiency bound} of $\go$.
Generalizing also the notion of complete passivity, $\go$
will be called a {\em completely semipassive}
state if all its finite tensorial powers are
semipassive with respect to one fixed efficiency bound $\E$.

Evidently, a state is completely
semipassive in all inertial frames if it is completely passive
in some inertial frame. The following theorem (Thm. 3.3 in \cite{Kuc01})
shows that if conversely,
a state is completely semipassive in a given inertial frame, then there
exists an inertial frame where it is completely passive.

\begin{theorem}\label{Hauptsatz}
The state $\go$ is completely semipassive with efficiency bound
$\E$ if and only if there exists a ${\bf u}\in\reals^s$ with
$|{\bf u}|\leq\E$ such that with respect to $H+{\bf uP}$,
$\go$ is a ground state or a KMS-state at a finite
inverse temperature $\beta\geq0$.
\end{theorem}

In a relativistic theory, the Hamiltonian of the system moving at
velocity ${\bf u}<c$ is not $H+{\bf u}{\bf P}$, but
$\gamma(H+{\bf u} {\bf P})$,
where $\gamma=(1-|{\bf u}|^2/c^2)^{-\frac{1}{2}}\equiv
(1-|{\bf u}|^2)^{-\frac{1}{2}}$. Theorem
\ref{Hauptsatz} still holds without any modification, but the
inverse temperature of the system is not the
parameter $\beta$ found there, but
$\beta/\gamma$.

\section{Semipassivity and $\beta$-boundedness}\label{semibeta}

Thm. \ref{Hauptsatz} 
describes a most generic example of a nonequilibrium
state, and the question is whether bounds on the power of a
cyclic process could be of interest in less generic situations
than that of a translation invariant state.
As far as such investigations are concerned, it is an obstacle of
the above definition of semipassivity that the invariance of
$\go$ is part of the definition and its motivation. While the
problem addressed in Thm. \ref{Hauptsatz} is nontrivial only if
$\go$ is invariant under all spacetime translations, it would be
of interest whether the semipassivity condition can be subdivided
into this invariance property plus some additional condition that
may be meaningful in other situations as well. Such a condition
can be obtained by modifying the following notion, which has first
been investigated by Guido and Longo:

\begin{definition}\label{betabound}
The state $\go$ is called {\bf $\beta$-bounded with bound 1}
if the linear space
$\dM\gO$ is a subspace of the domain of $e^{-\beta H}$ and
if the set $e^{-\beta H}\dM_1\gO$ consists of vectors with lengths $\leq1$
where $\dM_1:=\{A\in\dM:\,\|A\|\leq1\}$.

$\go$ is called {\bf completely $\beta$-bounded} if
for each $n\in\naturals$,
the state $\go^{\otimes n}$ on the algebra $\dM\otimes\dots\otimes\dM$
is $\beta$-bounded with bound 1.
\end{definition}

The following theorem shows that this notion characterizes
thermodynamic equilibrium states at a finite and nonnegative
inverse temperature $\beta$:

\begin{theorem}[Guido, Longo]\label{X}
$\go$ is completely $\beta$-bounded if and only if it is a ground state
or a KMS-state at an inverse temperature $\geq2\beta$.
\end{theorem}

Now modify Def. \ref{betabound} as follows:

\begin{definition}
The state $\go$ is called {\bf semi-$\beta$-bounded} if there
exists a damping factor $\E\geq0$ such that the linear space
$\dM\gO$ is a subspace of the domain of $e^{-\beta(H+\E|{\bf
P}|)}$ and the set $e^{-\beta(H+\E|{\bf P}|)}\dM_1\gO$ consists
of vectors with length $\leq1$.

It is called {\bf completely semi-$\beta$-bounded} if for each
$n\in\naturals$, the state $\go^{\otimes n}$ on the algebra
$\dM\otimes\dots\otimes\dM$ is semi-$\beta$-bounded with respect
to one fixed damping factor $\E\geq0$.
\end{definition}

One then obtains the following modification of Thm. \ref{X}
(Thm. 6 in \cite{Kuc01a}):

\begin{theorem}
A stationary and homogeneous state $\go$ is completely
semi-$\beta$-bounded with respect to a damping factor $\E\geq0$
if and only if there exists a ${\bf u}\in\reals^s$ with $|{\bf
u}|\leq\E$ such that $\go$ is a ground state or a KMS-state at an
inverse temperature $\geq 2\beta$ with respect to $H+{\bf uP}$.
\end{theorem}

As $|{\bf P}|$ is a positive operator, the operator
$e^{-\beta\E|{\bf P}|}$ is bounded and provides an additional
damping term, so $\beta$-boundedness implies
semi-$\beta$-boundedness for all $\E\geq0$, i.e.,
semi-$\beta$-boundedness is the weaker assumption.

\section{Passivity and vacuum states}\label{vacuum}

If $\go$ is a vacuum state, then the considerations of the Introduction
suggest that $\omega$ is passive with respect to
each Hamiltonian of the form
$\gamma(H+{\bf v}{\bf P})$, where
$|{\bf v}|<c=1$ and
$\gamma=(1-{\bf v}^2/c^2)^{-1/2}$.

In what follows, we assume this and, in addition, that
$\omega$ is a pure state. It has been shown in \cite{Kuc01}
that under these assumptions,
the joint spectrum of $H$ and ${\bf P}$ is contained in the cone
$$V_+:=\{(\eta,{\bf k})\in\Rd:\,\eta\geq0,\eta^2-{\bf k}^2\geq0\},$$
i.e., the spectrum condition holds.

To further proceed now,
we need some basic structures
of local quantum fields, which associate von Neumann algebras
$\dM(\O)$ of local observables with all bounded open spacetime regions
$\O\subset\Rd$ in such a way that the following conditions are satisfied:
\begin{itemize}
\item {\bf Isotony.} If $\O$ and $P$ are bounded open regions in $\Rd$
such that $\O\subset P$, then $\dM(\O)\subset\dM(P)$.

\item {\bf Locality.} If $\O$ and $P$ are spacelike separated bounded open
regions in $\Rd$ and if $A\in\dM(\O)$ and $B\in\dM(P)$, then $AB=BA$.

\item {\bf Spacetime Translation Covariance.} The
representation $V$ of \newline
$(\Rd,+)$ satisfies
$$V(x)\dM(\O)V(x)^*=\dM(\O+x)$$
for all bounded open sets $\O\subset\Rd$ and for all $x\in\Rd$.

\item {\bf Spectrum Condition.} The joint spectrum of the generators of
$V$ is contained in the closed forward light cone.

\item $\dM$ is assumed to be the smallest
von Neumann algebra that contains all local algebras $\dM(\O)$
associated with bounded open regions.
\end{itemize}

The trajectory of a (pointlike) observer who is uniformly accelerated
in the 1-direction with acceleration $a$ can be translated to the
curve
$$\tau\mapsto\frac{c^2}{a}\left(\sinh\frac{a\tau}{c},
\cosh\frac{a\tau}{c},0,\dots,0\right),$$
where $\tau\in\reals$
denotes the accelerated observer's eigentime. The wedge
$W_1:=\{x\in\Rd:\,x_1>|x_0|\}$, which is referred to as the {\em
Rindler wedge}, is the region of all spacetime points the accelerated
observer can communicate with using causal signals. Therefore, the
elements of the algebra $\dM(W_1)$ are precisely those observables the
uniformly accelerated observer can measure. The images of $W_1$ under
Poincar\'e transformations are referred to as {\em wedges}.

We assume that some uniformly accelerated observer exists:
\begin{itemize}
\item There is a self-adjoint operator
$K_1$ generating, within $W_1$, the free dynamics of the uniformly
accelerating observer, i.e., $$e^{i\tau K_1}\dM(\O)e^{-i\tau
K_1}=\dM(\Lambda_1(\mbox{$\frac{a}{c}$}\tau)\O)$$ for all $\tau\in\reals$ and
all bounded open sets $\O\subset W_1$.
$K_1$ strongly commutes with $P_2,\dots,P_s$, and $K_1\gO=0$.
\end{itemize}
Here, $\Lambda_1(\frac{a}{c}\tau)$ denotes the Lorentz boost by
$\frac{a}{c}\tau$ in the 1-direction.
$\dM$ is not yet assumed to be
covariant under a full representation of the Poincar\'e group,
although the assumption that $K_1$ strongly commutes with
$P_2,\dots,P_s$ is already a part of this condition. The following
result has been proved in \cite{Kuc01}.

\begin{proposition}
If the pure state $\go$ satisfies the spectrum condition and if
$\go$ exhibits passivity with respect
to the dynamics generated by $K_1$, then $\go$
is a KMS-state of $\dM(W_1)$ with respect to $K_1$ at
the Unruh temperature
$$T_U=\frac{\hbar a}{2\pi ck}.$$
\end{proposition}

\section{Conclusion}\label{conc}

The above results show that on the one hand, moving states of 
matter violate passivity, and for moving homogeneous states 
the extent to which this is the case can be considered as a 
kind of a ``distance'' from thermodynamic equilibrium.

On the other hand, vacuum states are characterized by the 
property that they are passive in the eyes of every 
observer whose motion does not enforce violation of passivity
due to nonstationary inertial forces. 

These two observations may help clarifying the notion of a 
particle, which is particularly fuzzy in quantum field theories
on curved spacetimes (already in the Rindler wedge). 
Particles should generate some kind of
``wind'' in the eyes of some observer, so this wind should be the
appropriate indicator whether or not they are present. 
With regard to the Unruh effect, this point of view confirms
that the vacuum state is, indeed, a state without
any particles, even in the accelerated observer's eyes.

\section*{Acknowledgements}
This work has been supported by the Stichting FOM, and 
the results reported here have been partly reported and 
partly obtained during the Conference on Operator Algebras
and Mathematical Physics 2001 at Constanta, where I profited
from discussions with D. Guido. They have been 
worked out at the Korteweg-de Vries Institute of Mathematics,
Amsterdam. I would like to thank M. Porrmann for carefully 
reading the manuscript.


\begin{thebibliography}{*****}

\bibitem{BW75} Bisognano, J. J., Wichmann, E. H.: On the duality Condition
for a Hermitian Scalar Field. {\it J. Math. Phys.} {\bf 16}, 985-1007 (1975)

\bibitem{BW76} Bisognano, J. J., Wichmann, E. H.: On the Duality Condition
for Quantum Fields. {\it J. Math. Phys.} {\bf 17}, 303-321 (1976)

\bibitem{Bor00} Borchers, H.-J.: On the Revolutionization of Quantum
  Field Theory by Tomita's Modular Theory. {\it J. Math. Phys.} {\bf
    41}, 3604-3673 (2000)

\bibitem{BB98} Borchers, H.-J., Buchholz, D.: Global Properties of Vacuum
States in de Sitter Space. {\it Ann. Poincar\'e Phys. Theor. A} {\bf 70},
23-40 (1999)

\bibitem{BB94} Bros, J., Buchholz, D.: Towards a Relativistic KMS-Condition.
{\it Nucl. Phys. B} {\bf 249}, 291-318 (1994)

\bibitem{BDFS98} Buchholz, D., Dreyer, O., Florig, M., Summers, S. J.:
Geometric Modular Action and Spacetime Symmetry Groups, {\it Rev. Math.
Phys.} {\bf 12}, 475-560 (2000)

\bibitem{BFS} Buchholz, D., Florig, M., Summers, S. J.: The Second
Law of Thermodynamics, TCP, and Einstein Causality in Anti-de Sitter 
Space-Time. {\it Class. Quant. Grav.} {\bf 17}, L31-L37 (2000)

\bibitem{BMS01} Buchholz, D., Mund, J., Summers, S. J.: Transplantation
of Local Nets and Geometric Modular Action on Robertson-Walker
Space-Times. {\it Fields Inst. Commun.} {\bf 30}, 65-81 (2001)

\bibitem{BuWi} Buchholz, D., Wichmann, E. H.: Causal Independence
and the Energy-Level Density of States in Local Quantum Field Theory,
{\it Commun. Math. Phys.} {\bf 106}, 321-344 (1986)

\bibitem{GL01} Guido, D., Longo, R.: Natural Energy Bounds in Quantum
Thermodynamics, {\it Commun. Math. Phys.} {\bf 218}, 513-536 (2001)

\bibitem{Haa92} Haag, R.: {\it Local Quantum Physics}. Berlin: Springer 1992

\bibitem{HHW} Haag, R., Hugenholtz, N. M., Winnink, M.: On the Equilibrium
States in Quantum Statistical Mechanics. {\it Commun. Math. Phys.} {\bf 5},
215-236 (1967)

\bibitem{HaSw} Haag, R. Swieca, J. A.: When Does a Quantum Field Theory 
Describe Particles, {\it Commun. Math. Phys.} {\bf 1}, 308-320 (1965)

\bibitem{Kas76} Kastler, D.: Equilibrium States of Matter and Operator
Algebras. {\it INAM Symposia Mathematica} {\bf XX}, 49-107 (1976)

\bibitem{Kubo} Kubo, R.: Statistical Mechanical Theory of Irreversible
Processes, I. {\it J. Math. Soc. Japan} {\bf 12}, 570 (1957)

\bibitem{Kuc00} Kuckert, B.: Two Uniqueness Results on the Unruh Effect
and on PCT-Symmetry. {\it Commun. Math. Phys.} {\bf 221}, 77-100 (2000)

\bibitem{Kuc01} Kuckert, B.: Covariant Thermodynamics of Quantum
Systems: Passivity, Semipassivity, and the Unruh Effect. Preprint
{\tt hep-th/0107236}, to appear in {\it Ann. Phys. (N. Y.)}

\bibitem{Kuc01a} Kuckert, B.: $\beta$-Boundedness, Semipassivity,
and the KMS-Condition. Preprint {\tt math-ph/0111037}

\bibitem{MarS} Martin, P. C., Schwinger, J.: Theory of
Many Particle Systems, I. {\it Phys. Rev.} {\bf 115}, 1342 (1959)

\bibitem{Mun01} Mund, J.: The Bisognano-Wichmann Theorem for Massive
Theories. Preprint {\tt hep-th/0101227}, to appear in {\it Ann H. Poincar\'e}

\bibitem{PW78} Pusz, W., Woronowicz, S. L.: Passive States and KMS States for
General Quantum Systems. {\it Commun. Math. Phys.} {\bf 58}, 273-290 (1978)

\bibitem{Unr76} Unruh, W. G.: Notes on black-hole evaporation. {\it Phys. Rev.}
{\bf D14}, 870-892 (1976)

\end{thebibliography}
\end{document}